# Enabling Bio-Plausible Multi-level STDP using CMOS Neurons with Dendrites and Bistable RRAMs

Xinyu Wu
Boise, ID, USA
tomas.wu@gmail.com

Vishal Saxena
Department of Electrical and Computer Engineering
University of Idaho
Moscow, ID, USA
vsaxena@uidaho.edu

*Abstract*— **Large-scale integration of emerging nanoscale non-volatile memory devices, e.g. resistive random-access memory (RRAM), can enable a new generation of neuromorphic computers that can solve a wide range of machine learning problems. Such hybrid CMOS-RRAM neuromorphic architectures will result in several orders of magnitude reduction in energy consumption at a very small form factor, and herald autonomous learning machines capable of self-adapting to their environment. However, the progress in this area has been impeded from the realization that the actual memory devices fall well short of their expected behavior. In this work, we discuss the challenges associated with these memory devices and their use in neuromorphic computing circuits, and propose pathways to overcome these limitations by introducing 'dendritic learning'.**

## I. Introduction

Human brain is capable of processing unstructured information sensed from the environment, and performing real-time pattern discovering and recognition tasks in a remarkably fast, accurate, robust, and energy-efficient manner. Thanks to a half century of advances in semiconductor technology, modern computers can perform such tasks but still require orders of magnitude higher energy, as well as specialized programming. Massive parallelism, event-driven spike-based communication and *in-situ* synaptic plasticity are believed to be responsible for brain's effective and energy-efficient information processing. Recently, brain-inspired neuromorphic hardware have demonstrated impressive ultra-low power performance in implementing convolutional neural networks [1]. However, it is not amenable to accommodate a huge number of synapses and cannot adjust the synaptic weights while in operation. In the past decade, the discovery of spike-timing-dependent-plasticity (STDP) and emerging of nanoscale resistive random-access memory (RRAM) devices has opened new avenues towards the realization of brain-inspired computing. Many RRAMs have demonstrated small feature size ($4F^2$), ultra-energy-efficiency (pJ/switch), CMOS compatible and 3D integration capability, and exhibit bio-plausible STDP characteristics [2], [3]. Studies also suggested STDP can be used to train spiking neural networks (SNNs) with RRAM synapses *in-situ* without trading-off their parallelism [4], [5].

Consequently, it is natural to envision hybrid CMOS-RRAM very-large-scale integrated (VLSI) circuits to achieve dense integration of CMOS neurons and RRAM synapses to build neural-inspired computing chips by leveraging the nanometer scale silicon processing technology. Recently, mixed-signal chips with spiking neurons that can interface with RRAM devices, and learning algorithms and circuits built using these neural motifs have been demonstrated by the authors [6]–[8].

A densely-integrated CMOS-RRAM spiking neural network, with non-volatile analog-like weights, is ideal for such realization [4], [7]. However, neuromorphic circuit community is facing challenges in practical realization of these implementations, where a major roadblock is the bistable and probabilistic switching behavior of RRAM [9]–[12]. A majority of small-size RRAMs exhibits abrupt switching nature, which in consequence limits stable synaptic resolution to 1-bit (or binary, bistable); furthermore, their switching probability and switching time are typically depends on the voltage applied to the device, as well as the duration of the voltage pulse. Fig. 1 illustrated the abrupt resistance decrease (SET) and the corresponding switching dependence on the voltage in a $HfO_x$ RRAM device [13]. To circumvent these issues, compound memristive synapse with multiple bistable devices in parallel was recently proposed to emulate analog weights [12]–[14]. Moreover, a standard pattern classification application has shown that least 4-bit of synaptic resolution is needed to achieve reasonable recognition performance [15]. However, simply placing multiple bistable RRAM in parallel doesn't result in the expected exponentially shaped learning window that appears in biological STDP, which has been found critical for guaranteeing computing stability and efficiency in theoretical analyses [16]–[18].

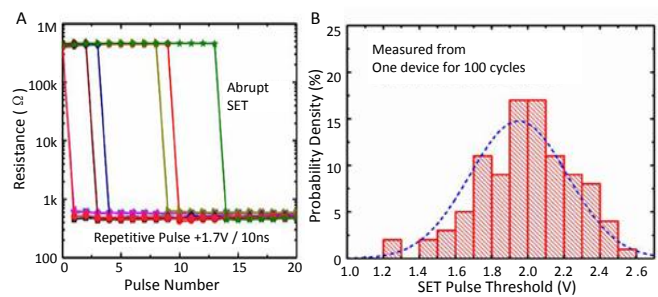

Fig.1. Binary stochastic switching and the STDP in RRAM. (A) Abrupt SET transition starting from the off-state by repetitive SET pulses. (B) Measured statistical distribution of pulse amplitude required for triggering the SET switching from the off-state (adapted from [13]).

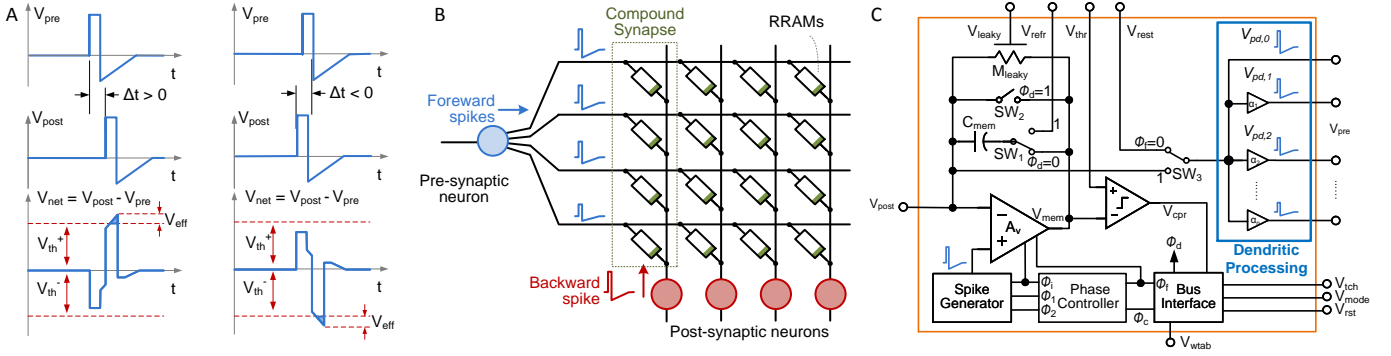

Fig.2. The proposed with CMOS neural network with dendritic processing. (A) A pair of spikes are applied across a synapse to create relative-timing dependent net potential $V_{net}$, of which the over-threshold portion ($V_{eff}$) may cause the RRAM resistance to switch. (B) A single layer of spiking neural network with RRAM synapses organized in crossbar architecture. Post-synaptic neuron connects with a pre-synaptic neuron with several RRAM synapses in parallel. Back-propagated spikes after dendritic processing modulate RRAM in-situ under STDP learning rule. (C) Schematic diagram of the proposed CMOS neuron with dendritic processing architecture. Parallel attenuators ($\alpha_i$) reduce voltage amplitude of backward spikes for parallel RRAM devices when CMOS soma fires. During non-fire (integration) mode, all parallel RRAM devices are connected to same current summing point.

The fundamental contribution of this work is that we propose a new concept where "dendritic processing" is added to the compound synapse. This is an important step where bistable switching devices, which have been practically realized, be researchers, can be employed in stochastic regime to realize synapses with multi-bit resolution and lead to practical realization of large-scale spiking neural networks on chip. Leveraging the bistable probabilistic switching, we specifically demonstrate that the proposed compound synapse with dendritic processing can realize high-resolution plasticity with the much desired exponentially-shaped STDP learning window.

## II. REALIZING MULTIBIT SYNAPSES USING BISTABLE RRAMS

In pair-wise STDP learning, spikes sent from pre- and post-synaptic have their voltage amplitudes below the positive and negative switching thresholds ($V_{th}^+$ and $V_{th}^-$) of a bipolar RRAM device. RRAM switching events may occur only if this spike pair overlaps and creates a net potential ($V_{net}$) greater than the switch threshold, as illustrated in Fig.2A. When several bistable RRAM devices are combined in parallel in a crossbar configuration to realize a compound synapse with multi-bit resolution, all the RRAMs in parallel will have a same switching probability as a function of applied voltage, if they are simply connected between two terminals.

To enable RRAM in the compound synapse with spike-timing-dependent switching probability, dendritic processing is applied to the pre-synaptic neuron output as shown in Fig. 3B. Here, a single-layer neural network is implemented in the crossbar architecture with a RRAM device, as an electrical synapse, connects pre- and post-synaptic neurons at each crosspoint. The original spike waveform from the pre-synaptic neuron runs through multiple dendritic branches before reaching the binary synaptic devices, and the spike amplitudes are reduced in these branches depending on their attributes. Since the synaptic devices' switching depends on the voltage and duration applied on their two terminals, these post-dendritic spike waveforms after dendritic processing produce several different voltage amplitudes and cause the synaptic devices to switch with different probabilities (assuming same pulse duration): the greater of the post-dendritic spike amplitude, the higher probability to switch; the lower of the post-dendritic spike amplitude, the less probability to switch. Thus, from statistical standpoint, the average conductance of the compound synapse with $n$ binary RRAM devices in parallel under a given voltage $V_i$ can be written in mathematical formula as

$$\overline{G_{cs}}(V) = \sum_{i=1}^{n} p_i(V_i) \frac{1}{R_{ON,i}} + \sum_{i=1}^{n} (1-p_i(V_i)) \frac{1}{R_{OFF,i}}, \quad (1)$$

where $p_i$ is the SET switching probability of the $i^{th}$ device, and $R_{ON}$ and $R_{OFF}$ are the RRAM resistance at ON and OFF state respectively. Generally, $R_{OFF}$ is greater than $R_{ON}$ by several orders of magnitude [19], and thus, can be neglected in overall conductance. For simplicity, assuming $R_{ON}$ of all devices is same [1] and normalizing it to one, the probability of the compound synapse equals the normalized conductance value $g$ is

$$p(\overline{G_{cs}}(V) = g) = \sum_{i=1}^{n} p_i(V_i). \quad (2)$$

Above formula shows the synergy effect of parallel RRAMs could approximate a nonlinear function if $p_i$ and $V_i$ are not linear at the same time. So, the proposed dendritic processing provides a new dimension to manage the amplitude and timing of spikes, the overall STDP learning curve now can be designed to a desired shape more easily. This is enabled by the key fact that the devices are operated in probabilistic switching regime, and each device switches differently with respect to the

---

[1] $R_{ON}$ of a real RRAM could be stochastic as well. See an example in [19].

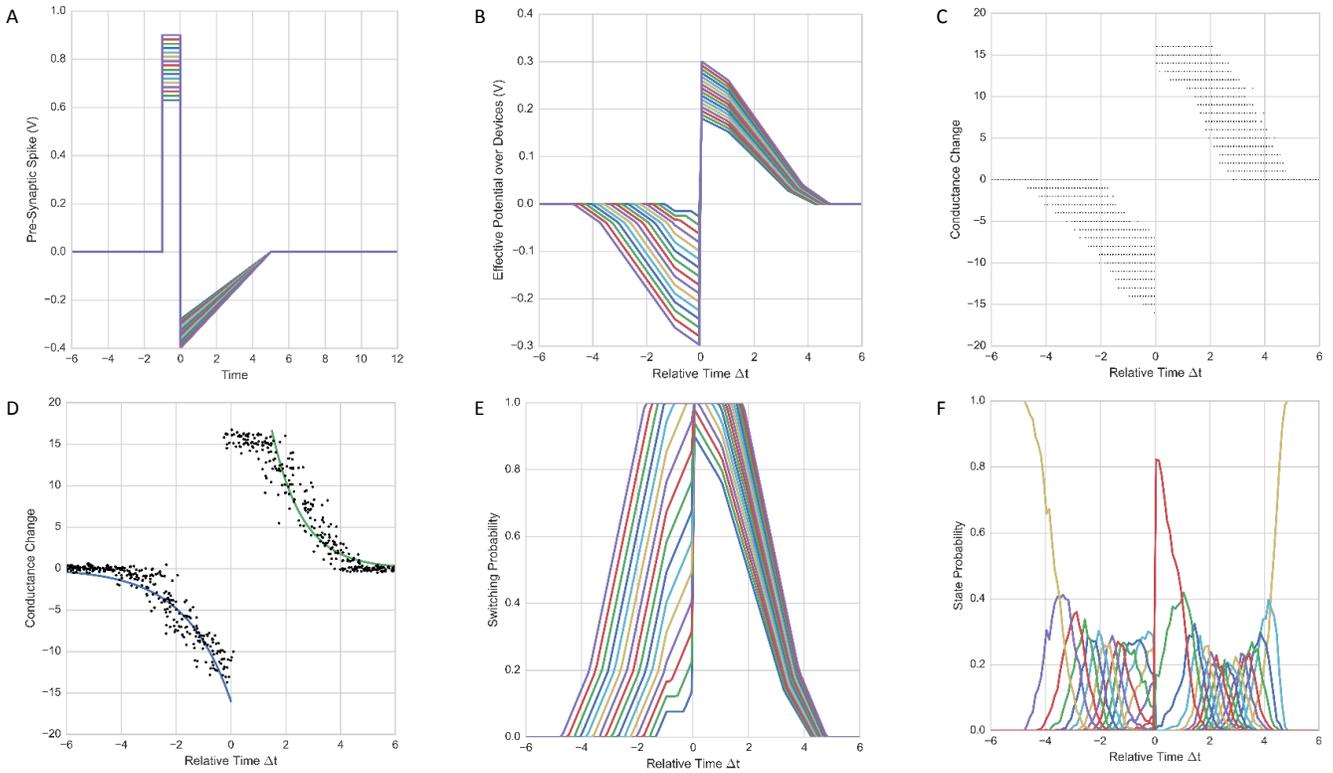

Fig. 3. STDP learning window created by compound binary resistive synapses with dendritic processing. (A) Pre-synaptic spike waveforms: Dendritic processing attenuates the voltage amplitude of spike with factors from 0.5 to 1, and creates 16 parallel spikes. (B) Effective potential $V_{eff}$ over parallel devices versus the relative arrival timing of post- and pre-synaptic spikes. 16 levels are created over $V_{th}^+$ and $V_{th}^-$. (C) The compound synapse shows an equivalent of 4-bit (16 levels) weight STDP leaning. Each dot represents a possible resistance value. (D) STDP window with dot density presenting the probability (with noise and jitter). The double exponential curves fit well to the maximum values of probability. (E) Switching probability for the 16 devices versus the relative pre/post timing (Δt). The asymmetric shape is created by dendritic processing on the pre-synaptic neuron only. (F) Probability as a function of relative timing of the compound synapse at each of its conductance state. The state probability curves are spaced with each other in an increasing exponential manner.

time difference between the pre- and post-spikes (*Δt*)

$$V_i = f(\Delta t). \quad (3)$$

In term of circuit realization, each dendritic branch is implemented by adding an attenuator to the CMOS soma output in a compact circuitry as shown in Fig. 3B. One of several possible realizations is a resistor ladder followed by source follower buffers. In detailed, noting the spike waveform generated by the spike generator as $V_a^+$ and $V_a^-$, then the dendritic processing generates *n* post- dendritic spikes

$$\begin{cases} V_{pd,i}^+ = \alpha_i V_a^+ \\ V_{pd,i}^- = \alpha_i V_a^- \end{cases}, \quad i = 1,2,\cdots,n \quad (4)$$

where $V_{pd,i}^+$ and $V_{pd,i}^-$ is the positive and negative amplitudes of the $i^{th}$ post- dendritic spike, and $\alpha_i$ is the attenuation factor of the $i^{th}$ dendritic branch.

The CMOS soma could be implemented in an integrate-and-fire circuits and winner-takes-all (WTA) mechanism in a reconfigurable architecture based on single opamp, as presented in [6]–[8]. In one configuration, the CMOS soma is designed to provide a constant voltage at the neuron's current summing input, and then, allows reliable and linear spatio-temporal spike integration by charging membrane capacitor with the flow-in currents converted by passive and resistive synaptic devices. In another configuration, the CMOS soma generates STDP- compatible spike and drives the spikes propagating in both forward and backward directions with high energy-efficiency. In the third configuration, this soma circuitry supports local learning with many neurons organized in a group and becomes selective to the input patterns through competition and lateral inhibition with a shared WTA bus [7].

III. EXPIRIMENTAL RESULTS

The architecture is implemented and simulated in Python 2.7, and 10,000 epochs were performed for stochastic simulation.

The stochastic switching of the RRAM synapse is modeled by the cumulative probability of a Gaussian distribution as experimentally demonstrated in [13]

$$p(V_{net}) = \int_0^{V_{net}} \frac{1}{\sqrt{2\sigma^2 \pi}} e^{\frac{-(x-V_{th})^2}{2\sigma^2}} dx, \quad (5)$$

where $p(V_{net})$ is the SET or RESET switching probability under net potential $V_{net}$ applied across the two terminals of the RRAM; $V_{th}$ is the average threshold voltage, and σ is the standard deviation of the distribution. Here, for demonstration purpose we choose $V_{th}^+ = 1V$ and $V_{th}^- = -1V$ for the positive and negative thresholds respectively, with σ = 0.1.

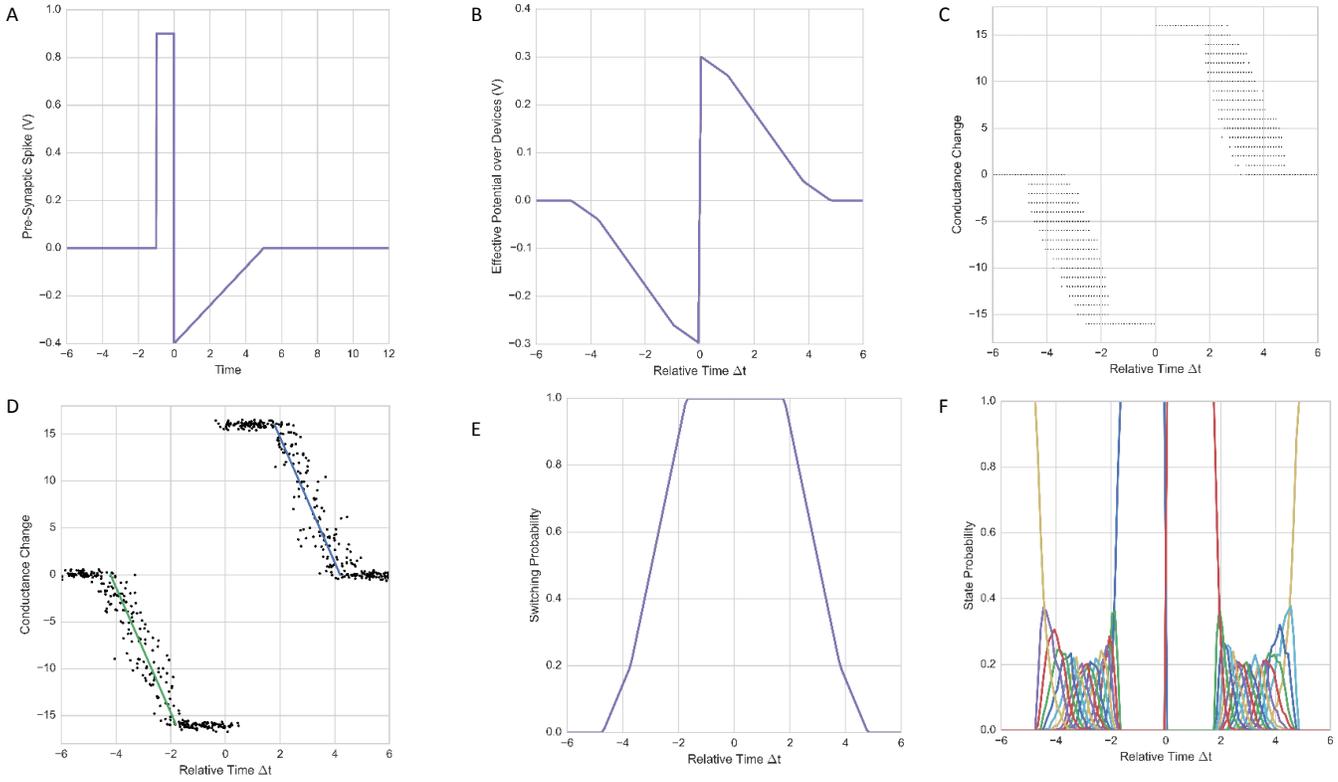

Fig. 4. STDP learning window created by compound binary resistive synapses without dendritic processing. (A) Pre-synaptic spike waveform. (B) Effective potential $V_{eff}$ over parallel devices versus the relative arrival timing of post- and pre-synaptic spikes. (C) Equivalent 4-bit STDP leaning window. (D) STDP window with dot density presenting the probability. The double linear curves fit to the maximum values of probability. (E) Switching probability of the 16 devices versus the Δt. (F) Probability of the compound synapse at each conductance state.

In simulation, a hardware-friendly spike waveform [4], [7] is selected. This spike waveform has a constant positive shape and a linearly rising negative tail, and has been demonstrated on a CMOS neuron chip by the authors [6]. With a normalized timescale, the positive tail has 0.9V amplitude and spans 1 time unit; the negative tail has a peak amplitude of 0.4V and spans 5 time units. Attenuating factors ($α_i$) of the dendritic attenuators were set to values linearly spanning from 0.6 to 1. Using these dendrites, 16 parallel spikes were generated as shown in Fig. 3A. These pre-synaptic spikes yield 16 positive and 16 negative levels in the STDP learning function that spans from -5 to 5 time units, as shown in Fig.3B.

Simulation results in Fig.3C shows the STDP learning windows with normalized conductance changes, $g$, where each dot represents a possible state of the synapse. It clearly demonstrates an equivalent of 4-bits weight with 16 levels both on the positive as well as negative sides of the STDP window. Using each dot to represent the probability density of the state, Fig.3D depicts the maximum likelihood state of the conductance change under STDP learning, and now a double exponential curve fits well to the simulated results. Here, each dot represents the state of the conductance compound synapse in the stochastic simulation. Mimicking the biological measurements, Gaussian noise and jitters were added to the synaptic change. The STDP learning shows a plateau when Δt in the range of 0 to 1 because the switching probabilities of the parallel devices are easier to saturate when only a portion of the pre-synaptic spike's tall positive tail overlaps with the post-synaptic spike's negative tail. Ideally, this plateau can be eliminated by using a very narrow positive tail for the spike waveform which corresponds to faster switching characteristics of the RRAM device.

Fig.3E further illustrates the asymmetry of the STDP curves, which plot the individual switching probability of the 16 devices versus Δt. We can observe that the RRAMs' switching probability curves are dense in the left-hand panel, due to the smaller amplitude of the spike's negative tail than its positive tail. Dense probability curves yield narrow span of their combined distribution, and are easier to saturate especially when the positive tail of pre-spike partially overlapping with the post-spike and creates a less change than full overlapping to the net potential respecting to the Δt. In this simulation, almost a half of the RRAM devices were saturated from Δt = 0 to 1. Another view is shown in Fig.3F, which depicts the probability of the compound synapse to occupy each of the normalized conductance state. It shows the state-wise probability curves are spaced with an increasing exponential manner; while a significant dominating state of

$g = 16$ when $\Delta t$ falls into the range of 0 to 1.

To provide a comparison of the proposed dendritic approach with a simple compound synapse, the simulation results of stochastic STDP learning without dendritic processing is illustrated in Fig. 4. Although this achieves 4-bit weight through STDP leaning, however, the STDP learning curves are linear as shown in Fig. 4D due to the same switching probability of all the devices in parallel as shown in Fig. 4E. And the state-wise probability curves are equally spaced as shown in Fig. 4F.

## IV. Conclusion

The proposed compound synapse with dendritic processing realizes exponential STDP learning similar to biology, while using practically feasible bistable memory devices. This can potentially enable ultra-low-power and significantly compact machine learning hardware for large-scale spiking neural networks that require plasticity with multibit resolution. Immediate applications will include practical realization of spike-based convolutional neural networks and restricted Boltzmann machines in a chip-scale form factor. This architectural exploration using the proposed compound probabilistic synapses can help benchmark the expected behavior from the emerging RRAM devices; nanoscale RRAM devices with large resistances will help realize lower power consumption.